# Loop State-preparation-and-measurement tomography of a two-qubit system

M. E. Feldman,[1] G. K. Juul,[1] S. J. van Enk,[2] M. Beck[1,*]

[1]*Department of Physics, Whitman College, Walla Walla, Washington 99362*
[2]*Department of Physics, University of Oregon, Eugene, Oregon 97403*
*Corresponding author: beckmk@whitman.edu*

**We experimentally demonstrate that loop state-preparation-and-measurement (SPAM) tomography is capable of detecting correlated errors in a two-qubit system. We prepare photon pairs in a state that approximates a Werner state, which may or may not be entangled. By performing measurements with multiple different detector settings we are able to detect correlated errors between two single-qubit measurements performed in different locations. No assumptions are made concerning either the state preparations or the measurements, other than that the dimensions of the states and the positive-operator-valued measures describing the detectors are known. The only other needed information is experimentally measured expectation values, which are analyzed for self-consistency. This demonstrates that loop SPAM tomography is a useful technique for detecting errors that would degrade the performance of multiple-qubit quantum information processors.**

## I. INTRODUCTION

Loop state-preparation-and-measurement (SPAM) tomography is a technique for detecting correlated errors between state preparations and/or measurements performed on a quantum system [1-4]. By correlated errors we mean the following. Suppose that Alice and Bob perform measurements on a two-qubit system, in order to verify that their particles violate a Bell inequality. To do this properly Alice's measurement settings must be independent of Bob's settings. However, it might be the case that their settings are not independent—some of Bob's settings might depend on (be correlated with) Alice's settings. In this case it is possible for the particles to *appear* to violate a Bell inequality, even if they are in a state that is not capable of violating this inequality [5-8]. This is one of the types of correlated errors that loop SPAM tomography is sensitive to.

One of the major advantages of loop SPAM tomography is that it relies on a minimum of assumptions. The only assumption needed is that we know the dimensions of the Hilbert spaces that describe the state of the system (via its density operator), and the measurements (via the positive-value-operator measures, POVMs, that describe the detectors). Other than that, all that is needed is experimentally measured expectation values. By analyzing an over-complete set of measurements, loop SPAM tomography looks for self-consistency within the data. Here self-consistency means that for each state preparation we can assign a single state, and for each measurement we can assign a single POVM. If the data is not self-consistent in this manner, we say that there are correlated errors. Hence, in the scenario described above, Alice and Bob can use loop SPAM tomography to detect the correlated errors between their measurements because they don't need to know the state of their particles, nor have any knowledge of what their measurement settings actually are.

Loop SPAM tomography has recently been demonstrated to be effective for detecting correlated errors between state preparations and measurements in a single-qubit system [4]. Here we show that it also works for detecting correlated errors between spatially separated measurements performed on a two-qubit system (as proposed in Ref. [3]).

To demonstrate the effectiveness of loop SPAM tomography for characterizing two-qubit systems we prepare photon pairs in states that approximate Werner states; these states are a mixture of a Bell state with a maximally mixed state (an incoherent background) [9]. The states we create have a sufficient degree of mixture to ensure they are not capable of violating the Clauser-Horne-Shimony-Holt (CHSH) inequality $S \leq 2$ [10, 11]. Correlated errors are introduced into the measurements, and with these errors the results appear to show that $S > 2$, which would violate the inequality. However, loop SPAM tomography is used to detect the presence of these correlated errors and show that the violation is erroneous.

One way that correlated errors could potentially be introduced into this system (and how they are described below) is as follows. Suppose that Alice and Bob wish to establish a secure communications channel by performing device independent quantum key distribution (QKD) [6, 12]. In such a scheme they can verify the security of their key by demonstrating that their particles violate a Bell inequality. However, an eavesdropper (Eve) manages to obtain prior information about Alice's and Bob's measurement settings, for example, by hacking into their computers. Eve then performs select rotations on their particles, effectively changing their measurements. She does this in a manner such that the apparent degree of correlation between the measurements is increased, yielding $S > 2$. This allows Eve to fool both Alice and Bob into believing that their key is secure, when in fact it is not. Finally, Alice and Bob suspect that Eve might be present, so they perform loop SPAM tomography on their data to detect the correlated errors that Eve introduces and to expose her.

While this might be an implausible scenario in practice, correlated errors could also arise from hardware or software problems that are unknown to Alice or Bob. The important point is that loop SPAM tomography works regardless of where the correlated errors come from. Beyond the particular experiments described here, our results demonstrate the ability to detect correlated errors in a device-independent way, with a minimum of assumptions, in a multiple-qubit system. As such, loop SPAM tomography is useful for characterizing quantum information processors. Fault-tolerant quantum computers are susceptible to correlated errors, and as the fidelity of quantum operations improve they will be sensitive to ever smaller errors. It will be important to have a means of detecting these small but correlated errors, in order to improve our trust in the results of quantum computations.

In Sec. 2 we present a brief description of the theory of loop SPAM tomography. We describe our experimental setup and present our results in Sec. 3, and in Sec. 4 we have some concluding remarks.

## 2. THEORY

### A. Loop SPAM Tomography

An unbiased detector that measures polarization is described by a POVM that has $n = 3$ independent parameters, and we can construct Hermitian operators (linear combinations of POVMs) that correspond to polarization measurements that also are described by three independent parameters [1, 4, 13]. Let Alice's measurement operators be denoted by $\hat{A}_i$, where $i$ labels the different measurement settings that Alice can choose. Similarly denote Bob's measurement operators by $\hat{B}_j$. If Alice and Bob have a source of photon pairs whose joint polarization state is described by the density operator $\hat{\rho}$, the expectation value for a joint measurement is given by

$$E_{ij} = \mathrm{Tr}\left[\hat{\rho}\left(\hat{A}_i \otimes \hat{B}_j\right)\right]. \quad (1)$$

Assume that $\hat{\rho}$ is held constant during the measurements, and consider the $E_{ij}$'s to be elements of a matrix $\overline{E}$; we use an overbar to denote a quantity expressed as a matrix. Each row of $\overline{E}$ refers to a fixed measurement for Alice $\hat{A}_i$, while each column refers to a fixed measurement for Bob $\hat{B}_j$. Because each POVM has 3 independent parameters, the largest matrix $\overline{E}$ that can consist entirely of independent elements is 3x3.

Consider the case where Alice and Bob make an over-complete set of measurements. Alice makes measurements with $2n = 6$ different detector settings and so does Bob. The 6x6 matrix of expectation values $\overline{E}$ can be partitioned into corners consisting of 3x3 matrices as follows

$$\overline{E} = \begin{pmatrix} \overline{A} & \overline{B} \\ \overline{C} & \overline{D} \end{pmatrix}. \quad (2)$$

The $n$x$n$ corner matrix $\overline{A}$ consists of independent measurements, but the other corners cannot be independent of $\overline{A}$. For example, matrix $\overline{A}$ is connected to matrix $\overline{B}$ in the sense that they share a common set of Alice's measurements, and the measured matrix elements of $\overline{B}$ must be consistent with that fact.

Define the partial determinant of $\overline{E}$ as $\Delta(\overline{E}) \equiv \overline{A}^{-1}\overline{B}\overline{D}^{-1}\overline{C}$. It is called the partial determinant because it is mathematically similar to the determinant, but it is not a scalar quantity—it is a matrix of smaller size than $\overline{E}$. It can be shown that the measured data are internally consistent as described above, and free of correlated SPAM errors, if and only if $\Delta(\overline{E}) = \overline{1}$, where $\overline{1}$ is the 3x3 identity matrix [1-4]. No knowledge of the state $\hat{\rho}$ or the detector operators is necessary to make this determination. Indeed, we don't even need to estimate the operators that describe the state or the detectors. All we need are the measured $E_{ij}$'s, and knowledge of the Hilbert-space dimensions of the state and the POVMs that describe the detectors.

If Eve has prior knowledge of Alice's and Bob's detector settings, and uses that knowledge to effectively change their detector settings to achieve a desired outcome, she will create correlated errors. Thus, to determine if Eve is present, we construct the matrix of expectation values $\overline{E}$ as given in Eq. (2), and then calculate the partial determinant $\Delta(\overline{E})$. If $\Delta(\overline{E}) - \overline{1} \neq 0$, to within the statistical errors of the measurements, then Eve's interference has been detected.

Furthermore, it is not necessary for Alice and Bob to use the full $2n = 6$ settings. They can each make $n + 1 = 4$ measurements, and embed these measurements into a 6x6 matrix $\overline{E}$ and test the condition $\Delta(\overline{E}) = \overline{1}$ [1, 3, 4]. This is done as follows. The first $(n+1)$ rows and $(n+1)$ columns are made up of the independent elements. Columns $(n+2)$ through $2n$ are copies of columns 2 through $n$, and rows $(n+2)$ through $2n$ are copies of rows 2 through $n$.

### B. Horodecki M parameter

We wish to determine whether the state that we create is capable of violating the CHSH inequality or not, and we use the method described in Ref. [11] to do this. Given a density operator $\hat{\rho}$, we can create the 3x3 matrix $\overline{T}$ whose elements are given by $T_{ij} = \mathrm{Tr}\left[\hat{\rho}\left(\hat{\sigma}_i \otimes \hat{\sigma}_j\right)\right]$, where the $\hat{\sigma}_i$'s are the Pauli matrices. The symmetric matrix $\overline{U}$ is given by $\overline{U} = \overline{T}^T\overline{T}$, where $\overline{T}^T$ is the transpose of $\overline{T}$, and the two largest, positive, eigenvalues of $\overline{U}$ are denoted by $u$ and $\tilde{u}$. We define the parameter $M(\hat{\rho})$ as $M(\hat{\rho}) \equiv u + \tilde{u}$; it was proven by Horodecki *et al* that $S_{\max} = 2\sqrt{M(\hat{\rho})}$. Hence, $\hat{\rho}$ is capable of violating the CHSH inequality if and only if $M(\hat{\rho}) > 1$ [11].

## 3. EXPERIMENTS

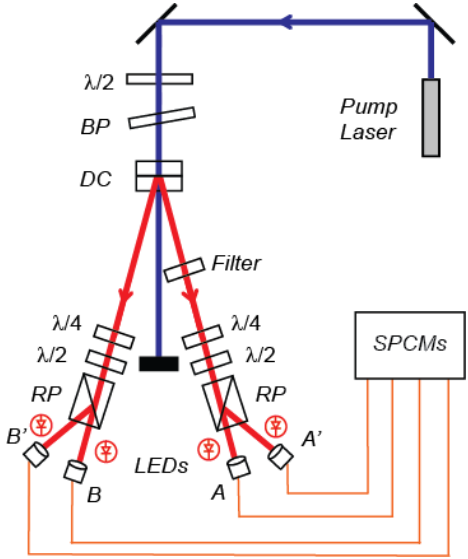

Fig. 1. The experimental configuration for performing two-qubit SPAM tomography. Here λ/2 denotes a half-wave plate, λ/4 denotes a quarter-wave plate, DC denotes down-conversion crystals, BP denotes a birefringent plate, RP denotes a Rochon polarizer, and SPCMs are single-photon-counting modules. The filter is an interference filter with a center wavelength of 810 nm, and a bandwidth of 10 nm.

**A. Experimental Design**

As shown in Fig. 1, we use a 150 mW, 405 nm laser diode to pump a pair of 0.5 mm thick beta-barium borate crystals, whose axes are oriented at right angles with respect to each other [14]. Each crystal produces Type-I spontaneous parametric down conversion, with signal and idler beams making angles of 3° from the pump. The relative amplitudes and phases of the two polarization components of the pump beam are controlled by a half-wave plate and a birefringent plate. Ideally the state that we are trying to produce is the Bell state

$$\left|\Phi^+\right\rangle = \frac{1}{\sqrt{2}}\left(\left|HH\right\rangle + \left|VV\right\rangle\right) . \quad (3)$$

However, our experimentally produced state is not pure. We model the state produced by our source as

$$\hat{\rho}_s(p_s) = p_s \left|\Phi^+\right\rangle\left\langle\Phi^+\right| + \frac{1-p_s}{2}\left(\left|HH\right\rangle\left\langle HH\right| + \left|VV\right\rangle\left\langle VV\right|\right). \quad (4)$$

This density operator represents our photons as being in the Bell state $\left|\Phi^+\right\rangle$ with probability $p_s$, and in an equal mixture of the states $\left|HH\right\rangle$ and $\left|VV\right\rangle$ with probability $1-p_s$. A state of this type is produced, for example, if there is some temporal walk-off between the wave packets for horizontal and vertical polarizations, which introduces a degree of distinguishability between them.

In order to add an additional degree of freedom to our state preparation procedure, we illuminate each of our detectors with light from a light-emitting diode (LED). Adjusting the current in each LED provides a controllable amount of unpolarized, incoherent background to the detected state, and we model this state as

$$\hat{\rho}_w(p_s, p_w) = p_w \hat{\rho}_s(p_s) + \frac{1-p_w}{4}\hat{1} . \quad (5)$$

This is a state that is in a mixture of state $\hat{\rho}_s(p_s)$ with probability $p_w$, and a state of purely random polarization with probability $1-p_w$. In the limit that $p_s = 1$ the states in Eq. (5) are Werner states, and we describe the states $\hat{\rho}_w(p_s, p_w)$ as approximating Werner states [9, 15]. The states $\hat{\rho}_w(p_s, p_w)$ have a Horodecki $M(\hat{\rho})$ parameter that is given by

$$M(\hat{\rho}_w) = p_w^2 + p_s^2 p_w^2 . \quad (6)$$

The expectation values in Eq. (1) are calculated as follows. Notice in Fig. 1 that Alice has two detectors: $A$ measures horizontally polarized photons while $A'$ measures vertically polarized photons (in combination with the wave plates in front of the Rochon polarizer these two detectors effectively measure orthogonal, elliptically-polarized photons). Bob has a similar pair of detectors. For a fixed setting of the wave plates, the joint probability that Alice measures a photon at $A$ and Bob measures a photon at $B$, $P(A,B)$, is given by the number of coincidence detections between $A$ and $B$, $N_{AB}$, divided by the total number of coincidence detections. In terms of the 4-sets of possible coincidences, this can be written as

$$P(A,B) = \frac{N_{AB}}{N_{AB} + N_{AB'} + N_{A'B} + N_{A'B'}} . \quad (7)$$

If both photons are measured to have the same polarization, we assign a value of 1 to the measurement, while if they have opposite polarizations we assign a value of –1. The expectation value for a fixed setting of the wave plates (fixed $i$ and $j$) is then given by the following combination of the 4 possible joint probabilities

$$E_{ij} = P(A,B) - P(A,B') - P(A',B) + P(A',B') . \quad (8)$$

Coincidence detections are measured with a commercial time-to-digital converter. We use a coincidence resolution of 3.2 ns.

**B. Experimental Results**

*1. No Background Illumination*

First we characterize our source with no background illumination from the LEDs. Initially Alice performs measurements with 2 different settings of her quarter- and half-wave plate angles [(0,0), (π/4,π/8)]; Bob also uses 2 different settings for his wave plates [(π/8,π/16), (-π/8,-π/16)]. Alice and Bob perform 10 trials of these measurements. From these measurements we can determine the CSHS parameter $S$, which is given by

$$S = E_{11} + E_{12} + E_{21} - E_{22} . \quad (9)$$

Using this expression we find a CHSH parameter of $S = 2.586 \pm 0.014$ (the uncertainty is the standard deviation of the 10 independent measurements), which violates local realism by 40 standard deviations and indicates that the source produces entangled-photon pairs.

Alice and Bob then perform additional measurements, as described in Ref. [16], in order to perform traditional quantum-state tomography (QST) on this source and to estimate its density operator $\hat{\rho}$. The resulting $\hat{\rho}$ is shown in Fig. 2 (a)-(b). We find that $\text{Tr}(\hat{\rho}^2) = 0.866$; this is a measure of the purity of the state. We then fit our measured $\hat{\rho}$ to the state $\hat{\rho}_w(p_s, p_w)$ of Eq. (5), using $p_s$ and $p_w$ as fit parameters. The best fit yields $p_s = 0.866$ and $p_w = 1.000$, which produces a fidelity between $\hat{\rho}_w(p_s, p_w)$ and $\hat{\rho}$ of $F = 0.985$. These parameters indicate that our source produces a state that is well described by Eq. (4). Using the density operator $\hat{\rho}$ that is returned from QST, and the method described in Sec. II.B, we find $M(\hat{\rho}) = 1.726$, which yields $S_{\max} = 2.628$. This is consistent with our directly measured value.

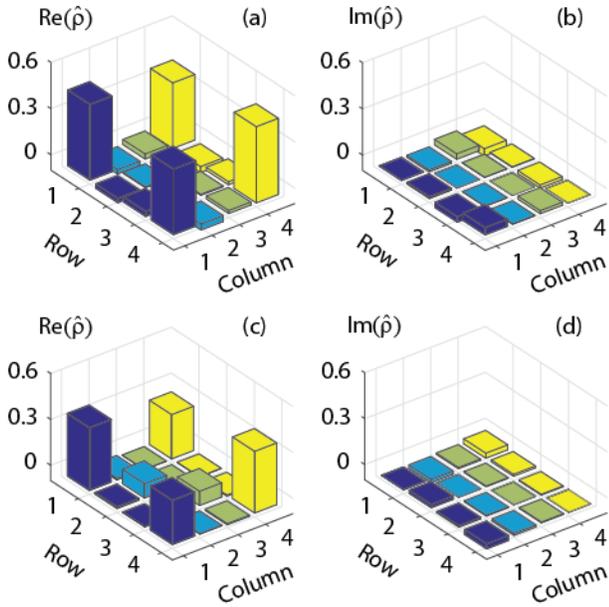

Fig. 2. The real and imaginary parts of the measured density matrices. The state created when the LEDs are off (no background illumination) is shown in (a) and (b), while the state created when the LEDs are on (background illumination) is shown in (c) and (d). For plotting these matrices we use the following ordering of the basis states: $|HH\rangle$, $|HV\rangle$, $|VH\rangle$, $|VV\rangle$.

*2. Background Illumination, No SPAM Correlations*

Next we turn on the LEDs to create an approximate Werner state, and perform measurements in which Eve is not present. Alice performs measurements with 4 different settings of her quarter- and half-wave plate angles [(0,0), (π/4,π/8), (π/4,0), (π/8,π/16)]; Bob also uses 4 different settings for his wave plates [(π/8,π/16), (-π/8,-π/16), (π/4,0), (π/4,π/8)]. After Alice and Bob complete one trial of these 16 measurements they repeat them; in total they perform 10 trials. We estimate $\hat{\rho}$ by performing QST on all the data, and the result is shown in Fig. 2 (c)-(d). The best fit $\hat{\rho}_w(p_s, p_w)$ has parameters $p_s = 0.928$ and $p_w = 0.628$, and a fidelity with $\hat{\rho}$ of $F = 0.997$. Furthermore, from this estimated $\hat{\rho}$ we find $M(\hat{\rho}) = 0.749$, which yields $S_{\max} = 1.731$; this indicates that the state is incapable of violating the CHSH inequality.

Note that the first 2 of Alice's and Bob's settings are those needed to determine $S$ directly from the measured data using Eq. (9). From these measurements we calculate $S = 1.710 \pm 0.017$. As expected, this does not violate the CHSH inequality.

Even though this state does not violate the CHSH inequality, it is still entangled. The negativity $N$ of $\hat{\rho}$ is defined as

$$N = 2\sum_k \max(0, -\lambda_k), \qquad (10)$$

where the $\lambda_k$'s are the eigenvalues of the partial transpose of $\hat{\rho}$ [17, 18]. For two-qubit states the negativity is 0 if the state is separable and 1 if it is maximally entangled. For the estimated state returned from QST we find $N = 0.397$. This value exceeds the lower bound on $N$ determined from our measured value of $S$ [19]

$$N \geq \frac{S}{\sqrt{2}} - 1 = 0.209. \qquad (11)$$

To perform loop SPAM tomography, the matrix elements $E_{ij}$ are determined directly from the measurement statistics [Eqs. (7) and (8)]. The 16 measurements of each trial are embedded in the 6x6 matrix $\bar{E}$ as described in Sec. 2.A, and the partial determinant $\Delta(\bar{E})$ is computed. The 10 trials allow us to determine the statistics of $\Delta(\bar{E})$. In Fig. 3(a)-(c) we show the mean and standard deviation of $\Delta(\bar{E}) - \bar{1}$, and the ratio of the absolute value of these two quantities. By examining the absolute value of the ratio of the mean to the standard deviation [Fig. 3(c)] we find that all of the matrix elements of $\Delta(\bar{E}) - \bar{1}$ are 0 to within one standard deviation, which indicates that no correlated SPAM errors were detected. This agrees with our expectations, as Eve is not present and hence should not introduce any correlated errors.

*3. Background Illumination, SPAM Correlations*

Now Alice and Bob make the same measurements on this same source, which is in the same state as described in Sec. 3.B.2. This state is incapable of violating the CHSH inequality, but Eve is now present. She uses prior knowledge of Alice's and Bob's measurement settings to rotate the polarizations of Bob's photons such that his settings are effectively transformed as follows [20]. When Alice's wave plates are set to (0,0) his first two measurements effectively become: (π/8,π/16) → (0,0), (-π/8,-π/16) → (0,0). When Alice's wave plates are set to (π/4,π/8) his first two measurements effectively become : (π/8,π/16) → (π/4,π/8), (-π/8,-π/16) → (-π/4,-π/8). Again, measurements are performed on 10 trials. With Eve's interference direct measurements of the CHSH parameter [Eq. (9)] yield $S = 2.447 \pm 0.014$, which appears to violate the CHSH inequality by 32 standard deviations.

However, since Alice and Bob are suspicious of Eve they perform loop SPAM tomography on the full set of data; the results are shown in Fig. 3(d)-(f). We find that several matrix elements of $\Delta(\bar{E}) - \bar{1}$ differ from 0 by over 6 standard deviations [Fig. 3(f)]. This indicates that correlated errors are present and that Eve's treachery has been detected.

## 4. DISCUSSION AND CONCLUSIONS

Above we described an experiment with a source that was incapable of properly violating the CHSH inequality, but appeared to violate it. This was possible because correlated errors increased the apparent degree of correlation between the measurements. However, loop SPAM tomography was used to detect these correlated errors, and demonstrate that the violation of the CHSH inequality was erroneous. This is possible because loop SPAM tomography works without information about either the source state or the measurements, so we don't have to rely on multiple assumptions about the source or the measurement devices. The only assumption we need is that the dimensions of the Hilbert spaces that describe the state of the source and the POVMs of the detectors are known. With this assumption, SPAM tomography is able to use only measured statistics to look for self-consistency within the data.

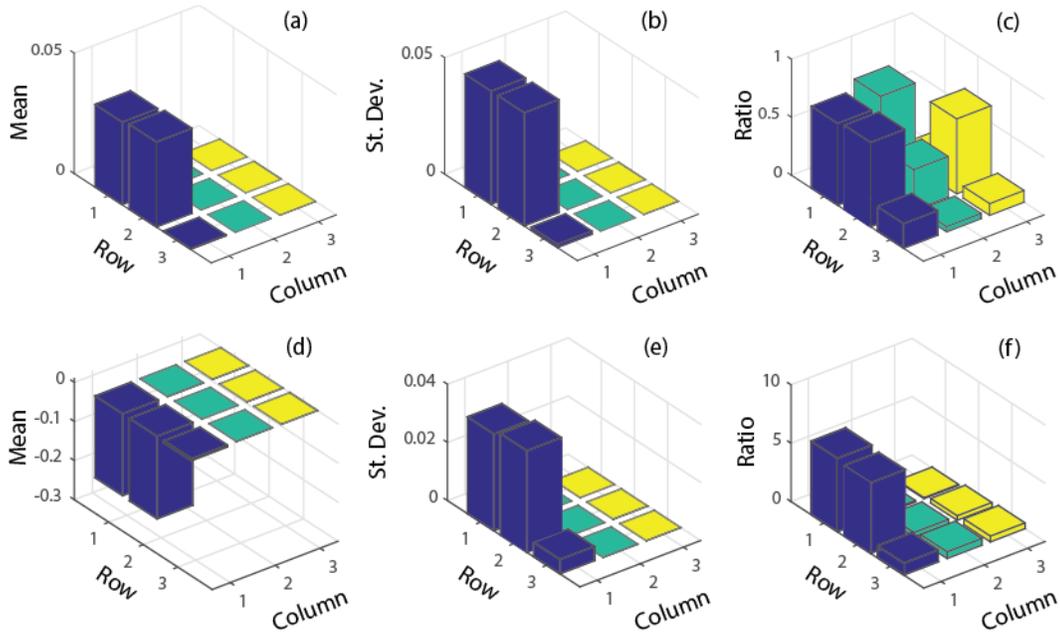

Fig. 3. (a) and (d) the mean of $\Delta(\bar{E}) - \bar{1}$, (b) and (e) the standard deviation of $\Delta(\bar{E}) - \bar{1}$, and (c) and (f) the absolute value of the ratio of these two quantities (mean divided by standard deviation). In (a)-(c) Eve is not present and there are no correlated errors. In (d)-(f) Eve introduces correlated errors.

While the errors were described above as if they originated from an eavesdropper, the actual source of errors is irrelevant. We have shown that loop SPAM tomography is capable of detecting correlated errors between measurements performed on two spatially separated qubits. This, combined with the fact that previous experiments have demonstrated its ability to detected correlated errors between state preparations and measurements [4], further demonstrates the utility of loop SPAM tomography. Because it works in a device-independent manner, with a minimum of assumptions, loop SPAM tomography promises to be a useful and effective tool for detecting correlated errors in a wide variety of quantum information processing systems.

**Funding Information.** National Science Foundation (NSF) (1719390); US Army Research Office (ARO) (W911NF-14-C-0048); Whitman College Louis B. Perry Summer Research Endowment and the Parents Student-Faculty Research fund.